\documentclass[usenatbib]{mn2e}
\usepackage{amssymb}
\usepackage{graphicx}
\usepackage{txfonts}

\newcommand{\integral}{{\textit{INTEGRAL}}}
\newcommand{\xte}{{\textit{RXTE}}}
\newcommand{\sax}{{\textit{Beppo\-SAX}}}

\newcommand{\fermi}{{\textit{Fermi GST}}}
\newcommand{\swift}{{\textit{SWIFT}}}
\newcommand{\g}{$\gamma$}
\newcommand{\ee}{e$^\pm$}
\newcommand{\msun}{{{\rm M}_{\sun}}}

\newcommand{\lsi}{LS~I~+61\degr 303}

\title[The TeV flare of Cygnus X-1]{A model of the TeV flare of Cygnus X-1: electron acceleration and extended pair cascades}

\author[A. A. Zdziarski, J. Malzac and W. Bednarek]
{Andrzej A. Zdziarski,$^1$\thanks{E-mail: aaz@camk.edu.pl (AAZ), 
malzac@cesr.fr (JM),  bednar@fizwe4.phys.uni.lodz.pl (WB)} Julien Malzac$^2$\footnotemark[1] and W. Bednarek$^3$\footnotemark[1]\\
$^1$Centrum Astronomiczne im.\ M. Kopernika, Bartycka 18, 00-716 Warszawa, Poland\\
$^2$CESR (Centre d'Etude Spatiale des Rayonnements), Universit\'e de Toulouse [UPS], CNRS [UMR 5187], 9 avenue du Colonel Roche, BP 44346,\\ 31028 Toulouse Cedex 4, France\\
$^3$Department of Astrophysics, University of {\L}{\'o}d{\'z}, Pomorska 149/153, 90-236 {\L}{\'o}d{\'z}, Poland\\
}

\date{Accepted 2008 November 25. Received 2008 September 18}

\pagerange{\pageref{firstpage}--\pageref{lastpage}}
\pubyear{2008}

\begin{document}

\maketitle

\label{firstpage}

\begin{abstract}
We consider theoretical models of emission of TeV photons by Cyg X-1 during a flare discovered by the MAGIC detector. We study acceleration of electrons to energies sufficient for TeV emission, and find the emission site is allowed to be close to the black hole. We then consider pair absorption in the photon field of the central X-ray source and a surrounding accretion disc, and find its optical depth is $\la 1$, allowing emission close to the black hole. On the other hand, the optical depth in the stellar field is $\sim$10 at $\sim$1 TeV. However, the optical depth drops with increasing energy, allowing a model with the initial energy of $\ga 3$ TeV, in which photons travel far away from the star, initiating a spatially extended pair cascade. This qualitatively explains the observed TeV spectrum, though still not its exact shape.  
\end{abstract}

\section{Introduction}
\label{intro}

The black-hole binary Cyg X-1 is at present the only X-ray binary that is both  certainly accretion-powered and from which very high energy \g-rays, $\ga 0.1$ TeV, have been detected by one of the current generation of TeV detectors. Namely, a transient TeV emission was detected by the MAGIC telescope (\citealt{albert07}, hereafter A07). Among the three other currently known TeV-emitting binaries (which are all persistent in the TeVs except for orbital variations), PSR B1259--63 \citep{aharonian05} is known to be powered by the rotation energy of a young pulsar rather than by accretion \citep{johnston92}, with the resulting interaction of the pulsar and stellar winds leading to production of energetic photons \citep{mt81}. Then, \lsi\ \citep{albert06} is very likely of the same nature, as indicated by its radio observations showing a cometary-tail--like variable structure rather than a jet \citep{dhawan06}. A possible further evidence for the presence of a young pulsar in that system is provided by a detection of a powerful $\sim$0.2-s flare with the peak 15--150 keV flux of $\simeq 5\times 10^{-8}$ erg cm$^{-2}$ s$^{-1}$ (higher by a factor of $\sim 10^3$ than the average source flux) and a blackbody-like spectrum by the \swift\/ BAT \citep{dp08,b08}. Such an event is consistent with a soft gamma-ray repeater/anomalous X-ray pulsar burst \citep{dg08}. Then, the case of LS 5039 \citep{aharonian06} is least clear, but the striking similarity of the broad-band spectra of all three systems points to similar mechanisms of their activity \citep{dubus06b}. All three persistent systems are also similar in being eccentric high-mass binaries containing massive B or O stars. Furthermore, their broad-band spectra are very much unlike those of accreting sources at similar Eddington ratios as well as are unlikely to be dominated by a collimated jet emission (see, e.g., a discussion in \citealt{z08}).

Then, there were some early claims of TeV emission from accreting X-ray binaries, most notably from Cyg X-3. Its TeV flux was claimed to be very strong and modulated at the orbital period (e.g., \citealt{lamb82}), which, however, was not confirmed by a subsequent Whipple Observatory pointing \citep{weekes83}. Cyg X-3 has not, until now, been detected by either the MAGIC, VERITAS, or HESS telescopes.

Thus, the case of TeV emission from Cyg X-1 is of crucial importance for our understanding of emission of \g-rays from accretion flows. The statistical significance of the detected transient was $4.9\sigma$ or $4.1\sigma$ post-trial (A07), unlikely to appear by chance. Supporting the reality of this transient emission is also the coincidence with a flare of Cyg X-1 seen in X-rays (A07; \citealt{t06,malzac08}, hereafter M08). 

The TeV detection occurred on 2006 September 24, MJD 54002.[928--982] (A07), which corresponds to the orbital phases of 0.90--0.91, i.e., with the black hole behind the companion. No signal was detected (A07) during the first half of that MAGIC observation, MJD 54002.[875--928], indicating time variability of the TeV emission on the time scale of less than an hour. The emission was detected in the 0.15--1 TeV range, and was fitted by a power law with a photon spectral index of $\Gamma=3.2\pm 0.6$ (A07). The 0.1--1 TeV isotropic luminosity corresponding to the best fit is $\simeq 2 \times 10^{34}$ erg s$^{-1}$ (assuming the distance of $D=2$ kpc, see \citealt{ziolkowski05} and references therein). The MAGIC telescope also observed, but did not detect, Cyg X-1 on 25 other nights, and obtained an upper limit to the flux vs.\ energy an order of magnitude below the flux during the flare (A07). 

As mentioned above, the TeV flare occurred during a longer X-ray outburst. Based on data from the \xte\/ All Sky Monitor, the X-ray flare lasted three days, about MJD 54002.0--54005.0. Cyg X-1 was then in the hard spectral state (M08). The \integral\/ IBIS and SPI instruments observed Cyg X-1 in the 18--700 keV range on MJD 54002.4--54004.3 (M08), during which the estimated bolometric luminosity was $\simeq 4 \times 10^{37}$ erg s$^{-1}$. M08 obtained, in particular, the X-ray spectrum for an interval coinciding with the TeV detection. The spatial location of the TeV emission region remains unknown. However, the simplest hypothesis to consider that it was close to the X-ray source. Then, we can calculate \ee\ pair absorption of the TeV photons as well as the Compton energy loss of accelerated electrons due to interactions with the X-rays.

\section{The binary parameters and the 1 \lowercase{e}V--1 M\lowercase{e}V spectrum}
\label{spectrum}

\begin{figure}
   \centering
   \includegraphics[width=0.7\columnwidth]{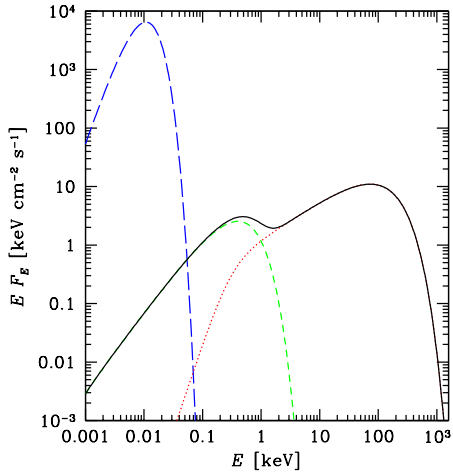}
      \caption{The spectral components of the Cyg X-1/HDE~226868 system. The long dashes give the blackbody spectrum of the OB supergiant (for the parameters of \citealt{herrero95}). The dotted curve gives the thermal Comptonization component from the hot inner flow. The short dashes give the spectrum of the surrounding optically-thick disc (tidally truncated at the outer radius), and the solid curve gives the spectrum of the disc together with the hot-plasma emission. See Section \ref{spectrum} for details.}
         \label{fig:spectrum}
   \end{figure}

The system parameters of interest to our study are the stellar temperature, $T_\star$, the stellar radius, $R_\star$, the separation, $a$, and the inclination, $i$. The radius of the companion, HDE 226868, and the component masses are subject to a considerable uncertainty, compare, e.g., \citet{herrero95} with \citet{ziolkowski05}. Here, we adopt the best model of \citet{herrero95}, with $R_\star=1.2\times 10^{12}$ cm, the separation of $a= 2.47 R_\star$, and $T_\star\simeq 3.2\times 10^4$ K. The observed stellar flux is plotted in Fig.\ \ref{fig:spectrum}. The value of the inclination of Cyg X-1 is relatively uncertain; here, we adopt $i=30\degr$ (e.g., \citealt{gb86}).

The 18--700 keV X-ray spectrum during the TeV flare was fitted by M08 as an e-folded power law. Physically, it corresponds to thermal Comptonization of blackbody photons produced at the inner edge of a cold disc. At low energies, the disc spectrum has a low-energy break, with the Rayleigh-Jeans form below the break. To model it, we assume the observed broad-band energy flux, $F_{\rm h}$, is given by,
\begin{equation}
E F_{\rm h}(E) = {K (E_{\rm b}/1\,{\rm keV})^{2-\Gamma} \exp(-E/E_{\rm c})\over 
(E/E_{\rm b})^{-3} + (E/E_{\rm b})^{\Gamma-2}},
\label{inner}
\end{equation}
which has the form of $K E^{2-\Gamma} \exp(-E/E_{\rm c})$ at $E\gg E_{\rm b}$. We assume $E_{\rm b}=0.3$ keV \citep{frontera01}. For the remaining parameters, we use the best-fit parameters of M08, $\Gamma=1.34$, $E_{\rm c}=111$ keV, $K=1.25$ keV cm$^{-2}$ s$^{-1}$. This spectrum is plotted in Fig.\ \ref{fig:spectrum}. 

The corresponding photon density within the central hot source, $n_{\rm h}(E)$, can be calculated by assuming an isotropic and homogeneous spherical source with the radius, $R_{\rm h}$, in which the average photon escape time is $3R_{\rm h}/4c$,
\begin{equation}
E^2 n_{\rm h}(E, R_{\rm h})\simeq {9 D^2 E F_{\rm h}(E)\over 4 R_{\rm h}^2 c}.
\label{conversion}
\end{equation}
The hard state emission is well modelled as coming from a hot inner flow surrounded by a cold disc truncated at $\sim$30--$100 R_{\rm g}$, where $R_{\rm g}$ is the gravitational radius (e.g., \citealt{dgk07}). For the black hole mass of $M_{\rm X}\sim 10\msun$, the radius of the inner hot flow is $R_{\rm h}\sim 10^8$ cm, which value we assume hereafter.

Moving away from the hot plasma, the local photon density includes an increasing number of soft photons from the surrounding disc. For a given height above the disc, most of the photons come from the part of the disc within the radius equal to the height. The inner disc temperature is approximately given by $2 kT_{\rm bb,max}\sim E_{\rm b}\simeq 0.3$ keV, with which factor the spectrum is exponentially cut off at high energies. The temperature decreases with the increasing radius as $T_{\rm bb}\propto R^{-3/4}$, and the resulting disc blackbody spectrum is $F_{\rm d}(E)\propto E^{1/3}$. At the outer edge of a considered region, $R_{\rm out}$, the spectrum changes to the Rayleigh-Jeans one. The corresponding energy is $E_{\rm out}=(R_{\rm out}/R_{\rm h})^{-3/4} E_{\rm b}$. Thus, we approximate the disc spectrum by,
\begin{equation}
E F_{\rm d}(E, R_{\rm out}) = {K' E_{\rm out}^{4/3} \exp(-E/E_{\rm b})\over 
(E/E_{\rm out})^{-3} + (E/E_{\rm out})^{-4/3}},
\label{outer}
\end{equation}
where the normalization constant, $K'$, follows from the condition that $F_{\rm d}$ and $F_{\rm h}$ intersect at some energy, which we choose as 1 keV based on the fits to \sax\/ data in Frontera et al.\ (2001). This choice also reproduces 
the soft excess appearing in the hard state of Cyg X-1 at $E\la 3$ keV \citep{gierlinski97}. We assume the disc is tidally truncated at $10^{12}$ cm. The spectrum of equation (\ref{outer}) for $R_{\rm out}=10^{12}$ cm and the total spectrum of the disc and the hot flow are plotted in Fig.\ \ref{fig:spectrum}. The disc photon density, $n_{\rm d}(E, R_{\rm out})$, is approximated analogously to equation (\ref{conversion}), 
\begin{equation}
E^2 n_{\rm d}(E, R_{\rm out})\simeq 9 D^2 E F_{\rm d}(E, R_{\rm out}) / (4R_{\rm out}^2 c).
\label{disc}
\end{equation}

\section{Electron acceleration}
\label{acceleration}

Here, we consider whether electrons can be accelerated to TeV energies within or close to the central X-ray source. We consider the acceleration rate in a generic form, applicable, e.g., to shock acceleration,
\begin{equation}
\dot \gamma_{\rm acc} ={\xi c \gamma\over r_{\rm L}}={\xi e B\over m_{\rm e} c}\simeq 1.8\times 10^7 \xi {B \over 1\,{\rm G}}\,{\rm s}^{-1},
\label{acc_rate}
\end{equation}
where $\gamma$ is the Lorentz factor, $R_{\rm L}=\gamma m_{\rm e} c^2/e B\simeq 1700\gamma/B$ cm is the Larmor radius, $B$ is the magnetic field strength, $e$ is the electron charge, and $\xi\la 1$ is a dimensionless quantity parameterizing the acceleration efficiency (see, e.g., \citealt{bg07}\footnote{Note that there is a spurious factor of $\upi$ in eq.\ (3) of \citet{bg07}, and the $-2$ factor in their eq.\ (6) should be outside the argument of the logarithm. }). The maximum electron Lorentz factor can be then obtained for the acceleration rate being equal to an energy loss rate. 

\begin{figure}
   \centering
   \includegraphics[width=0.7\columnwidth]{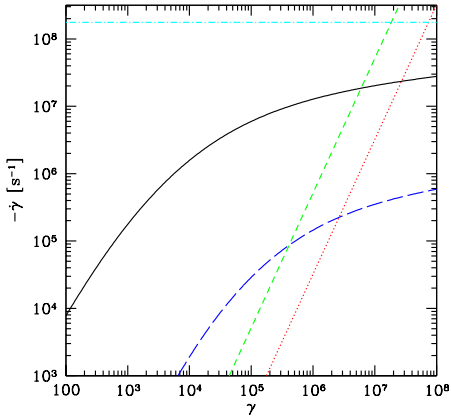}
      \caption{The Compton energy loss rate, $-\dot \gamma$, in the X-ray photon field, given by equations (\ref{inner}--\ref{disc}) and assuming the radius of $10^9$ cm (solid curve), and in the stellar field diluted at the position of the black hole (long dashes). The synchrotron loss rate is shown for $B=20$ G (short dashes), and for $B=5$ G (dots), and the dot-dashed line is ($-$) the acceleration rate, equation (\ref{acc_rate}) for $B=20$ G and $\xi=0.5$.
}
         \label{fig:gdot}
   \end{figure}

The synchrotron energy loss is given by, 
\begin{equation}
-\dot\gamma_{\rm S}={4\over 3}{\sigma_{\rm T}\over m_{\rm e} c} {B^2\over 8\upi}\gamma^2\simeq 1.3\times 10^{-9}\gamma^2 \left(B\over 1\,{\rm G}\right)^{2}\,{\rm s}^{-1}.
\label{gsyn}
\end{equation}
Equating this loss rate to the acceleration rate, equation (\ref{acc_rate}), we obtain the corresponding maximum Lorentz factor of
\begin{equation}
\gamma_{\rm max, S}=\left( 6\upi\xi e\over \sigma_{\rm T} B\right)^{1/2}\simeq 1.2\times 10^{8}\xi^{1/2} \left(B\over 1\,{\rm G}\right)^{-1/2}.
\label{gmaxs}
\end{equation}
Note that this implies a universal value of the maximum energy of the synchrotron component, as pointed out by \citet{gfr83} (and later by \citealt{dj96}). Taking the characteristic synchrotron energy of $(3/4\upi)\gamma^2 (h e B/m_{\rm e} c)$, we obtain,
\begin{equation}
E_{\rm max, S}={9\xi\over 16} {m_{\rm e} c^2\over \alpha_{\rm f}}\simeq 40 \xi\,{\rm MeV}
\label{emaxs}
\end{equation}
(where $\alpha_{\rm f}$ is the fine-structure constant), which is well below the TeV range. 

In addition, if electrons are accelerated within the hot plasma region, the Larmor radius has to be smaller than the acceleration site size, $R_{\rm acc}$, which yields the maximum Lorentz factor of
\begin{equation}
\gamma_{\rm max, L}={B R_{\rm acc} e\over m_{\rm e} c^2}\simeq 6\times 10^{5} {B\over 1\,{\rm G}} {R_{\rm acc}\over 10^9\,{\rm cm}}.
\label{gmaxl}
\end{equation}
Then, comparing equations (\ref{gmaxs}) and (\ref{gmaxl}), we obtain the optimum magnetic field yielding the highest maximum the Lorentz factor subject to both constraints together,
\begin{equation}
B_{\rm opt}=\left( 6\upi \xi m_{\rm e}^2 c^4\over R_{\rm acc}^2 e \sigma_{\rm T}\right)^{1/3} \simeq 35 \xi^{1/3} \left( R_{\rm acc}\over 10^9\,{\rm cm}\right)^{-2/3} {\rm G},
\label{bmax}
\end{equation}
which Lorentz factor equals
\begin{equation}
\gamma_{\rm max}=\left( 6\upi e^2 \xi R_{\rm acc}\over m_{\rm e} c^2  \sigma_{\rm T}\right)^{1/3} \simeq 2\times 10^7 \xi^{1/3} \left( R_{\rm acc}\over 10^9\,{\rm cm}\right)^{1/3}.
\label{gmax0}
\end{equation}
We see that the optimum magnetic field is relatively weak, e.g., $\simeq 160\xi^{1/3}$ G if $R_{\rm acc}=R_{\rm h}=10^8$ cm. The magnetic field expected in the hot inner plasma from equipartition arguments is considerably higher, $B\ga 10^5$ G (e.g., \citealt{mb08}). Thus, the preferred acceleration site is outside of the hot inner plasma. It can be, e.g., a shock region where the stellar wind collides with the hot inner flow, or an inner part of the jet. The Lorentz factor of the accelerated electrons is completely sufficient to account for the photon energies observed in the TeV flare. In fact, the maximum Lorentz factor can still be lower than that of equation (\ref{gmax0}), which will then allow a value of the magnetic field higher than that of equation (\ref{bmax}). 

An important process is also the loss rate due to Compton scattering with the X-ray and disc photons, equations (\ref{inner}--\ref{disc}), and due to the stellar photon density at the position of the black hole. For simplicity, we assume isotropic photon fields. The energy loss of electrons with $\gamma^2\gg 1$ in scattering with isotropic photons of an arbitrary energy is given by \citet{jones68} (see also eq.\ A30 in \citealt{z88}). We integrate this rate over the photon density at the radius of $10^9$ cm, obtaining the $-\dot\gamma$ shown in Fig.\ \ref{fig:gdot}, which curvature is due to to Klein-Nishina effects.We see that if the magnetic field is given by equation (\ref{bmax}), the Compton energy losses rate are less than those due to the synchrotron losses. Thus, they would only slightly affect our estimates above. Also, we see that the losses on the X-rays dominate over those on the stellar photons. 

Thus, we find the conditions around the X-ray source to allow electron acceleration up to $\gamma \sim 10^7$, which then easily allows emission of photons in the TeV range. The acceleration time scale, $\gamma/\dot\gamma_{\rm acc}$, is then very short, $\la 1$ s, see equation (\ref{acc_rate}).

\section{Pair absorption of T\lowercase{e}V emission}
\label{TeV}

Pair absorption of the TeV photons in photon-photon \ee\ pair production events is a major issue in interpretation of the TeV flare. In a study of pair absorption by photons from the companion star, \citet{bg07} found that TeV photons should be preferentially observed from Cyg X-1 at the orbital phases of 0.45--0.55, i.e., near the inferior conjunction, which is almost exactly opposite to the phase of the observed event. However, pair absorption can also take place on size scales much smaller than the system separation, due to pair-production events with the intense X-ray field of the central source. Below, we consider separately interaction with the X-rays and with the stellar photons.

\subsection{TeV--X-ray pair absorption}
\label{xray}

\begin{figure}
   \centering
   \includegraphics[width=0.65\columnwidth]{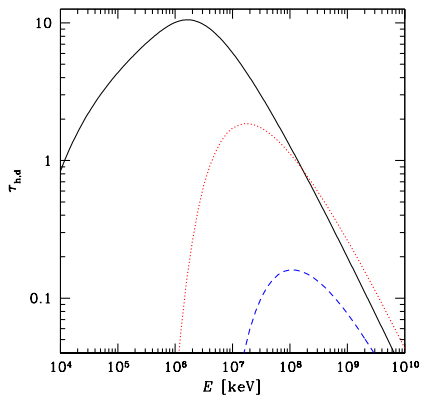}
      \caption{The optical depth to photon-photon pair production in the field of the hot plasma at $R=10^{8}$ cm (dolid curve), and of the disc emission. The 
dotted and dashed curves give the optical depth due to the disc photons along the disc axis from the height of $10^8$ and $10^9$ cm, respectively. The assumed inner disc radius is $10^8$ cm and the inner temperature is $kT_{\rm bb,max}=0.15$ keV. The acceleration site is likely at $\ga 10^9$ cm, Section \ref{acceleration}, and thus the TeV photons are only weakly attenuated in the field of the disc and the hot plasma.
}
\label{fig:tauX}
   \end{figure}

As discussed above, the acceleration site is most likely larger than the hot inner plasma. Still, for completeness we calculate the \ee\ pair absorption optical depth across the inner hot plasma, with the size of $R_{\rm h}$, 
\begin{equation}
\tau_{\rm h}(E_\gamma, R_{\rm h}) =R_{\rm h} \int_{m_{\rm e}^2 c^4/E_\gamma}^\infty\!\!\!\! n_{\rm h}(E) \bar\sigma_{\gamma\gamma}[(E_\gamma E)^{1/2}] {\rm d}E \propto R_{\rm h}^{-1}, 
\label{tauh}
\end{equation}
where $\bar\sigma_{\gamma\gamma}$ is the cross section averaged over isotropic target photons (\citealt{gs67}; corrected in \citealt{bmg73}; see also eqs.\ B11--B13 in \citealt{z88}). The resulting optical depth is shown in Fig.\ \ref{fig:tauX}. We see that whereas it is $\ga 1$ in the $\sim$0.01--100 GeV range, the (Klein-Nishina) high-energy decline of the cross section results in $\tau_{\rm h} \la 1$ above this range. 

We then consider pair absorption due to the disc photons. We follow the method of \citet{bednarek93}. The disc blackbody spectrum is taken as having the temperature of $kT_{\rm bb,max}=0.15$ keV at the disc inner edge of $R_{\rm h}=10^8$ cm, and diluted (due to a colour correction; the normalization is also affected by the exact value of $R_{\rm h}$, which is unknown) to match the level of the disc spectrum shown in Fig.\ \ref{fig:spectrum}. We calculate the optical depth along the disc axis from the height of $10^8$ and $10^9$ cm, see Fig.\ \ref{fig:tauX}. We see that even in the former case, it is $\la 1$ at $E\ga 0.1$ TeV. For the distance of $10^9$ cm (see Section \ref{acceleration}), pair absorption of photons $>0.1$ TeV on either the disc or hot-plasma photons is negligible. 

\subsection{Pair absorption on stellar photons and spatially extended pair cascades}
\label{star}

\begin{figure}
   \centering
   \includegraphics[width=5.3cm]{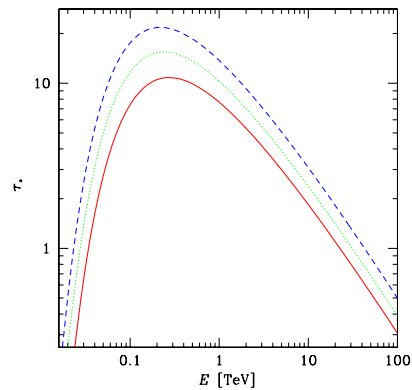}
      \caption{The pair-production optical depth in the field of the star from the location of the black hole up to infinity at the orbital phase of 0.9 and for the inclinations of $30\degr$, $45\degr$ and $60\degr$, from bottom to top.
}
         \label{fig:taubb}
   \end{figure}

We then calculate the optical depth on stellar photons, taking into account their anisotropy outside the stellar surface. We use the methods of \citet{bednarek97} (with triple integration) and \citet{dubus06a} (with quadruple integration) and found they fully agree with each other. The results for the orbital phase of 0.9, the binary parameters given in Section \ref{spectrum}, and three values of $i$ are shown in Fig.\ \ref{fig:taubb}. We see that at $i=30\degr$, $\tau_* > 1$ between 40 GeV and 20 TeV. This might have implied that the acceleration process could not have taken place in the black hole vicinity.

However, we note that a similar problem is faced in the case of the TeV-emitting X-ray binary LS 5039. Its parameters, $R_*\simeq 6\times 10^{11}$ cm, $T_*\simeq 3.9\times 10^4$ K, $a\simeq 2.2 R_*$, and the eccentricity of $e\simeq 0.35$ \citep{casares05}, imply a system even more compact and, especially at periastron, more opaque to pair absorption than  Cyg X-1. Still, a steady TeV emission is detected from LS 5039 at both the superior conjunction and the periastron \citep{aharonian06}, which two points occur very close from each other in the orbital phase in that system. This is contrary to the theoretical predictions based on the pair absorption alone, giving fluxes at those phases much lower than those observed \citep{dubus06a,bednarek06}, similarly to the predictions implied by Fig.\ \ref{fig:taubb}. The solution to this dilemma may be presented by pair cascades. They may be initiated by electrons and photons at energies of $\sim$3--10 TeV, at which energy the pair optical depth measured from the compact object drops to $\sim$1, implying that those photons can propagate relatively far from the star. The resulting pairs produce photons at lower energies, which then can propagate to the observer in the much diluted blackbody field. Such spatially extended pair cascade model was applied to LS 5039 by \citet{bednarek07}. Similar pair cascades can also take place in the X-ray field (see above). 
   
   \begin{figure}
   \centering
   \includegraphics[width=0.9\columnwidth]{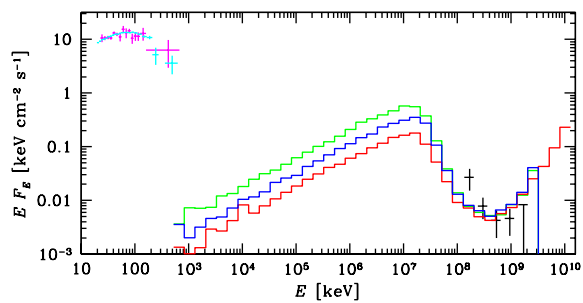}
      \caption{The \integral\/ spectrum from the ISGRI (cyan crosses), PICsIT (the cyan cross at the highest energy) and the SPI (magenta crosses), from M08, simultaneous with the MAGIC spectrum (black crosses, A07). The MAGIC spectrum is compared to the results from our pair cascade models, with the green and histograms corresponding to the primary monoenergetic electron injection at 3.16 TeV and 10 TeV, respectively. The blue histogram is for 3.16 TeV, but with the stellar temperature assumed to be $3.0\times 10^4$ K (which is within the observational uncertainties for Cyg X-1). The normalization of the histograms have been chosen to match the MAGIC data. }
         \label{fig:model}
   \end{figure}

To calculate pair cascades, we use two methods. One is of \citet{bednarek97}, which uses an approximation in the Klein-Nishina regime, and the other is an independent Monte Carlo method using the full Klein-Nishina cross section. We have found the two methods give results consistent with each other. We present some of our results of the latter in Fig.\ \ref{fig:model}, for monoenergetic electron injection at 3.16 TeV and 10 TeV. This form of injection is supported by results of studies of acceleration including energy losses, see, e.g., \citet{p04}. We find we can reproduce the MAGIC spectrum only qualitatively. Generally, our models predict too much emission at energies $\ga 1$ TeV, and too little at $\la 0.2$ TeV. We have found this to be a generic problem, and changing the injection energy does not improve the agreement with the data. Some improvement at the low energy end appears when a steep power law injection is added; however, this is at the expense of greatly increasing the intrinsic power of the flare. On the other hand, we consider it possible that, given complications in interpreting Cherenkov telescope data, the actual spectrum of Cyg X-1 was similar to that predicted by our models. At $\la 10$ GeV, the pair cascade process stops, and the spectrum has the $\Gamma=1.5$ photon index, as expected for an injection at high energies. The 0.1--1 TeV isotropic power is 0.011--0.018 of the the total power injected in TeV electrons, for the three models shown in Fig.\ \ref{fig:model}. This implies an injected power of $\sim 10^{36}$ erg s$^{-1}$, which is still much less than the bolometric X-ray luminosity, $\simeq 4\times 10^{37}$ erg s$^{-1}$ (M08). Note that most of the cascade power is emitted around $\sim$10 GeV, i.e., within the band of \fermi. However, its sensitivity\footnote{www-glast.slac.stanford.edu/software/IS/glast\_lat\_performance.htm} allows to detect such a flare only for a $\ga$30-hr duration. On the other hand, if there is a persistent emission at a level a factor of 10 below that of the flare, it would be detected by \fermi\/ within one year. 

In our model, we have assumed that the Compton losses of the cascade pairs dominate over the synchrotron losses. This requires that the magnetic field of the star is relatively weak, $B\la 5$ G or so (see Fig.\ \ref{fig:gdot}) along the line of sight. The magnetic field of HDE 226868 is unknown. In our model, it is required to be weaker than that of strongly-magnetized O stars, where the surface dipole field can reach $\sim 10^3$ G \citep{w06}. This is also pointed out by \citet{bka08}. We also assume that the pairs of the cascade quickly isotropise after their production, and produce \g-rays via Compton scattering locally, which will be the case even for a very weak field.

An alternative explanation of the observed flare is that the primary TeV emission occurred at a large distance from the black hole, e.g., along its jet \citep{pb08,bk08}, far enough to avoid the pair absorption by the stellar photons (see also \citealt{bka08}). The observed TeV variability (A07) implies the emission region has a size $\la 10 a$. The distance from the black hole needs to be of the order of a few times $a$ (fig.\ 2 in \citealt{bg07}). Even if it is $\sim 1a$, it corresponds to $\sim 2\times 10^6 R_{\rm g}$ for $M= 10\msun$. In the case of blazars, most of the \g-ray emission is considered to come from distances a few orders of magnitude smaller (e.g., \citealt{sbr94}). On the other hand, this has been challenged in some cases, especially for M87, where the TeV emission has been argued to originate at $\ga 120$ pc \citep{chs07}. For the black hole mass of $3\times 10^9\msun$, this corresponds to $\sim 3\times 10^5 R_{\rm g}$, still substantially less than corresponding number for Cyg X-1. 

Note that some energy is transported to such distances in Cyg X-1 in its jet, and it appears as radio emission \citep{stirling01}. However, the total radio luminosity of Cyg X-1 is only $\sim 10^{31}$ erg s$^{-1}$ \citep{fender00}, less than three orders of magnitude than the observed 0.1--1 TeV power, and five orders of magnitude less then the intrinsic cascade power, see above. 

\section{Conclusions}

We have studied opacity and acceleration models for the TeV flare observed from Cyg X-1 by the MAGIC telescope. We have found that spatially extended pair cascades allow some TeV photons to escape pair absorption on the stellar photons. This can explain the observations provided the companion star in the system has a weak magnetic field.

\section*{ACKNOWLEDGMENTS}

We thank G. Dubus, M. Ostrowski and {\L}. Stawarz for valuable discussions, and the referee for valuable suggestions. This research has been supported in part by the CNRS, the LEA Astrophysics Poland-France (Astro-PF) program, the Polish MNiSW grants NN203065933 and NN203390834, and the Polish Astroparticle Network 621/E-78/SN-0068/2007. We thank the MAGIC collaboration for the Cyg X-1 spectrum in electronic form.

\label{lastpage}
\end{document}